\documentclass[12pt]{iopart}
\usepackage{graphicx}
\usepackage[caption=false]{subfig}
\usepackage{epstopdf}
\usepackage{amsfonts}
    
\begin{document}

   \title{Proper-time measurement in accelerated relativistic systems}

   \author{Uri Ben-Ya'acov}

   \address{School of Engineering, Kinneret Academic College on
   the Sea of Galilee, \\   D.N. Emek Ha'Yarden 15132, Israel}

   \ead{uriby@kinneret.ac.il}

\vskip 2.0cm

\begin{abstract}
Separate constituents of extended systems measure proper-times on different world-lines. Relating and comparing proper-time measurements along any two such world-lines requires that common simultaneity be possible, which in turn implies that the system is linearly-rigidly moving so that momentary rest frames are identifiable at any stage of the system's journey in space-time.

Once momentary rest-frames have been identified, clocks moving on separate world-lines are synchronizable by light-signal communication. The synchronization relations for two clocks are explicitly computed using light-signals exchanged between them. Implications for the clock hypothesis are included. Also, since simultaneity is frame-dependent, incorrect usage of it leads to pseudo-paradoxes. Counter-examples are discussed.
\end{abstract}

\noindent{\it Keywords\/} : {proper-time; extended relativistic systems; relativistic rigid motion; rapidity; clock hypothesis; relativistic age.}

\vskip30pt

\section{Introduction}\label{sec: intro}

In classical Newtonian mechanics any set of particles may be grouped to form a ``system''. Time is absolute, independent of the referenc-frame, therefore common to all the chosen constituents, and a centre-of-mass (CM) may be unambiguously defined.

Relativity theory is closer to reality than Newtonian mechanics, telling us that time measurement is reference-frame-dependent. Then, except for the trivial case of a system whose constituents all move inertially with the same velocity, there is no common time for the system as a whole.

For point-like particles, proper-time is measured along their world-line, which may also be used for their age. Our originating research question is then {\it to what extent it is possible to assign the concept of common or characteristic proper-time to spatially extended systems}; in particular, with the wish to use this common proper-time for the {\it age} of the system.

For a system whose constituents all move inertially with the same velocity, the common proper-time is identified with the time reading of the common rest-frame. But if the system is not inertial the issue becomes far from trivial.

If the system is closed so that its total energy-momentum is constant and the CM frame of the system is inertial, then the common proper-time may be discussed in relation to dynamical quantities. This was done recently, to some extent, in refs.\cite{Internaltime2006,Inttimedil2007}. Here we wish to consider kinematically the case of non-inertial systems.

For a point-like particle moving on the world-line $x^\mu = \left(t,\vec r(t) \right)$ the proper-time lapse between two events, say A and B, is computed, as is well-known \cite{MTW73,Rindler2006}, by the integral
\begin{equation}\label{eq: Dtau}
 \Delta \tau_{\rm AB} = \int_{t_{\rm A}}^{t_{\rm B}} {\sqrt {dt^2 - {d\vec r}\,^2}} = \int_{t_{\rm A}}^{t_{\rm B}} {\sqrt {1 - {\vec v}\,^2} dt}
\end{equation}
If the particle is inertial this is the reading of a clock attached to the particle's rest-frame. Otherwise, this is the cumulative reading of clocks relative to which the particle is momentarily at rest.

Extended systems may be regarded as being composed of a number of point-like constituents. If the system accelerates then different constituents, moving on distinct world-lines, measure the proper-time differently. Then, in order to approach the issue of assigning a common proper-time to the whole system, the issue of {\it relating and comparing proper-times measured at distinct constituents} must be sorted out first. This is the purpose of the present talk.

Proper-time lapses measured along separate world-lines may be easily compared if the world-lines intersect twice, but then only between the intersections. This is the case with various versions of the so called `twins paradox'. Otherwise, in the absence of intersections, relating and comparing proper-time measurements at any two points of an extended system requires that some kind of simultaneity be possible. Therefore, to mimic in such cases the proper-time measurement for an inertial system, it follows that {\it momentary rest frames, common to the whole system, must be identifiable at any stage of the system's journey in space-time}. The last statement in italics is recognized as characterizing rectilinear relativistic rigid motion \cite{Born1909,HerglotzNoether1910}, which is possible for arbitrary (also time-dependent) accelerations (taking into account necessary differential accelerations between different points). Therefore, linearly-rigidly accelerated systems are used in the following to study comparative proper-time measurement in extended systems.

Since proper-times are Lorentz invariant quantities they should be treated in a Lorentz covariant manner. Linear relativistic rigid motion with general (not-necessarily constant) accelerations is discussed in the following Lorentz covariantly, allowing to relate accelerations, velocities and proper-times of arbitrarily different points along the moving system.

Simultaneity is then used to link and compare the time evolution of different parts of the system. For accelerating systems, the rapidity $\eta = \tanh^{-1}(v)$ \cite{RhodesSemon2004} is a very convenient parameter to identify the momentary rest-frames. Once simultaneity has been thus defined, clocks moving on separate world-lines are synchronizable by communicating light-signals, either between them or emitted from a source in between the clocks. The synchronization relations are computed and found to depend on the rapidity difference between emission and reception of the signals and the spatial separation of the clocks, and much less significantly on the details of the acceleration. These results are then used to support the relativistic clock hypothesis.

When momentary rest-frames are required, simultaneity can only be relative to the system as a whole, and should be referred to, accordingly, as {\it proper simultaneity}. Proper simultaneity is possible only for rigidly moving systems, and, since simultaneity is frame-dependent, incorrect application of it leads to wrong conclusions and appearance of so-called `paradoxes'. To emphasize this aspect of non-inertial motion, the article is finalized with demonstration of the ambiguity of proper-time comparison in non-rigid motion and discussion of two pseudo-paradoxes, Bell's spaceships `paradox' \cite{BellSSP} and Boughn's `identically accelerated twins' \cite{Boughn1989}.

{\it Notation}. The entire article is confined to Special Relativity only, referring to events $x^\mu = \left(x^0,x^1,x^2,x^3\right)$ in flat Minkowski space-time. With the convention $c=1$ and metric tensor $g_{\mu\nu} = {\rm diag} \left(-1,1,1,1\right) \, , \, \mu,\nu = 0,1,2,3$, for any 4-vectors $a^\mu = (a^0,\vec a)$ and $b^\mu = (b^0,\vec b)$ the inner product is $a \cdot b = -a^0 b^0 + \vec a \cdot \vec b$.

\vskip20pt

\section{Time measurement on spatially extended systems}\label{sec: timemeas}

Composite systems consist of a number (small or large) of points, each point moving on a separate world-line in space-time. At any such point, a virtual clock may be placed. Synchronization of separated clocks requires that simultaneity be established between them. Simultaneity is defined relative to a particular reference frame. If the clocks are inertial and relatively at rest then synchronization is naturally carried out in their common rest frame. Otherwise, for non-inertial clocks, common momentary rest frames identifying common momentary simultaneity hyper-planes must be found. Common momentary rest-frames are characteristic of linear rigid motion. We now review the notion of simultaneity, then consider accelerated linear motion and introduce into it the rigidity condition.

\subsection{Simultaneity and rigidity}\label{sec: simrig}

Let $x_{\rm A}^\mu \left( \theta_{\rm A} \right)$ and $x_{\rm B}^\mu \left( \theta_{\rm B} \right)$ designate the world-lines of two separate point-like entities, A and B, with $\theta_{\rm A}$ and $\theta_{\rm B}$ general time-like evolution parameters. Simultaneity is established between the two world-lines when there is a space-like displacement vector orthogonal to both : For each event $x_{\rm A}^\mu \left( \theta_1 \right)$ on A's world-line there is a unique event $x_{\rm B}^\mu \left( \theta_2 \right)$ on B's world-line so that the displacement vector $\Delta^\mu \left( \theta_1,\theta_2 \right) \equiv x_{\rm A}^\mu \left( \theta_1 \right) - x_{\rm B}^\mu \left( \theta_2 \right)$ is orthogonal to $x_{\rm A}^\mu \left( \theta_{\rm A} \right)$ at $\theta_{\rm A} = \theta_1$. In general $\Delta^\mu$ is not orthogonal to $x_{\rm B}^\mu \left( \theta_{\rm B} \right)$ at $\theta_{\rm B} = \theta_2$. Only when $\Delta^\mu$ is orthogonal to both $x_{\rm A}^\mu \left( \theta_{\rm A} \right)$ and $x_{\rm B}^\mu \left( \theta_{\rm B} \right)$ with the conditions
\begin{equation}\label{eq: simcond}
\Delta \left( \theta_1,\theta_2 \right) \cdot \frac{dx_{\rm A}}{d\theta_{\rm A}} \left(\theta_1\right) = 0  \quad \& \quad  \Delta \left( \theta_1,\theta_2 \right) \cdot \frac{dx_{\rm B}}{d\theta_{\rm B}} \left( \theta_2 \right) = 0
\end{equation}
may A and B be regarded {\it simultaneous}. If the simultaneity condition \eref{eq: simcond} is continuously maintained along these world-lines then $\Delta^\mu \left( \theta_1,\theta_2 \right)$ is of constant length\footnote{It is emphasized that the constancy of the distance $\Delta$ is due to the special relativistic context; this wouldn't necessarily be the case in a general rela1ivistic context.} $\sqrt{\Delta\cdot\Delta}$ and the motion is necessarily rigid.

The existence of a common rest frame, even momentarily, is a requisite : If two clocks are relatively moving, so they do not have a common rest frame, then there is no clear definition of simultaneity, even if the clocks themselves are inertial. Choosing different reference frames, in particular the rest frames of each clock, to determine simultaneity, determines different ratios between the time scales measured by the clocks (see \Sref{sec: nonrig}).

\vskip20pt

\subsection{Proper-time measurement in rectilinear accelerated motion}\label{sec: PTmeas}

The world-line of a point particle moving along the $x$-direction may be written, relative to some inertial reference frame S, as
 \begin{equation}\label{eq: xgen}
 x^\mu = \left( t,\xi^1(t),\xi^2,\xi^3 \right)
 \end{equation}
with constant $\xi^2,\xi^3$. Its unit 4-velocity is $u^\mu = \gamma(v) \left( 1,v,0,0 \right)$, with $v(t) = d\xi^1/dt$ and $\gamma(v) = \left( 1-v^2 \right)^{-1/2} = dt / d\tau$. The acceleration 4-vector at the point is
 \begin{equation}\label{eq: acc}
 a^\mu  = \frac{du^\mu }{d\tau} = \gamma^4 \left( v \right) \frac{dv}{dt} \left( v,1,0,0 \right) = a n_1^\mu \, ,
 \end{equation}
with $a = \gamma^2(v) \left( dv/d\tau \right)$ the proper acceleration and $n_1^\mu = \gamma(v)\left( v,1,0,0 \right)$ is the space-like unit 4-vector orthogonal to $u^\mu$ indicating the spatial direction of motion.

For linearly accelerating bodies it is very convenient to use the {\it rapidity} $\eta(v) \equiv \tanh^{-1}(v)$ -- the additive quantity in the superposition of co-linear velocities \cite{RhodesSemon2004} -- as the evolution parameter. It satisfies $d\eta = \gamma^2(v) dv = a d\tau$, so that the basic relation between the proper acceleration, the proper-time and the rapidity
 \begin{equation}\label{eq: tauaeta}
 a  = \frac{d\eta }{d\tau} \, ,
 \end{equation}
which holds for all rectilinear motion, is obtained \cite{Minguzzi2005,EJP2016a}.

In terms of the rapidity, the particle's unit 4-velocity and the spatial unit vector in the direction of motion are, respectively,
 \begin{equation} \label{eq: un1}
 u^\mu(\eta) = \left(\cosh\eta,\sinh\eta,0,0\right)  \quad , \quad  n_1^\mu(\eta) = \left(\sinh\eta,\cosh\eta,0,0\right) \, .
 \end{equation}
The world-line $x^\mu(\eta)$ then satisfies
 \begin{equation} \label{eq: xeta}
 \frac{dx^\mu}{d\eta}(\eta) = \frac{u^\mu(\eta)}{a(\eta)}
 \end{equation}
which upon integration yields
\begin{equation}\label{eq: intx}
 x^\mu(\eta) = \left(\int\limits^\eta {\frac{\cosh\eta}{a(\eta)} d\eta} \, , \, \int\limits^\eta {\frac{\sinh\eta}{a(\eta)} d\eta} ,0,0 \right) \, .
\end{equation}

It should be noted that under Lorentz transformations in the direction of motion the rapidity changes by an additive constant. Therefore, while the proper acceleration and rapidity differences are Lorentz invariant, $\eta$ as a variable is not, so the function $a(\eta)$ is necessarily frame-dependent.

\vskip20pt

\subsection{Rectilinear rigid motion} \label{sec: rectirig}

We now recall the essentials of rectilinear rigid motion \cite{EJP2016a}. Born's rigidity condition \cite{Born1909} implies that in rectilinear rigid motion there is always, continuously, a momentary rest frame that is common to all the constituents of the system. The Herglotz-Noether theorem \cite{HerglotzNoether1910} then verifies that such motion is possible with arbitrary accelerations.

The existence always of momentary rest frames common to the whole system implies that all its constituents move with the same (varying) spatial velocity $v$ and rapidity $\eta = \tanh^{-1}(v)$ relative to any inertial reference frame. Hence, the foregoing analysis in \sref{sec: PTmeas} is equally valid to all points and $\eta$ may be used as a common evolution parameter for the whole system.

In order to describe and analyze the motion of the system in space-time, some reference point must be initially chosen within it. This point defines a reference world-line $x^\mu = x_o^\mu(\eta)$ with $\tau_o(\eta)$ its proper-time. Following the foregoing discussion an orthonormal tetrad $\left\{ u^\mu ,n_i^\mu\right\}$ is chosen that is carried along the reference world-line, with $u^\mu(\eta)$ and $n_1^\mu(\eta)$ given by \eref{eq: un1} and the other two (constant) unit vectors $n_i^\mu \; (i=2,3)$ corresponding to displacements perpendicular to the spatial direction of motion. The tetrad $\left\{ u^\mu ,n_i^\mu\right\}$, corresponding to the common momentary rest-frames, is common to the whole system. The motion of the whole system is therefore completely determined by that of the reference world-line and the spatial triad $\left\{n_i^\mu (\eta)\right\}$.

The triad $\,\left\{ n_i^\mu (\eta), \, i = 1,2,3 \,\right\}$ spans the $\eta$-simultaneity hyperplanes relative to $x_o^\mu (\eta)$. These are the simultaneity hyperplanes relative to which synchronization of the clocks of the system is possible. Since all the points of the system share the same simultaneity hyperplanes, the choice of the reference point is arbitrary.

Any other point in the system may be defined relative to the reference world-line by a set of 3 constant distance parameters $\left\{\zeta^i\right\}$ relative to the triad $\left\{n_i^\mu (\eta)\right\}$, with the world-line
 \begin{equation} \label{eq: xxi}
 x^\mu (\zeta,\eta) = x_o^\mu (\eta) + \zeta^i n_i^\mu (\eta)
 \end{equation}
Let $a_o(\eta)$ be the proper acceleration of the reference point. From \eref{eq: xeta} and \eref{eq: xxi} then follows
 \begin{equation} \label{eq: uAtauo}
 \frac{d}{d\eta} x^\mu(\zeta,\eta) = \left(\frac{1}{a_o} + \zeta^1 \right) u^\mu(\eta) \, .
 \end{equation}
Since all unit 4-velocities are parallel they must be identical, $u^\mu(\zeta,\eta) = u^\mu(\eta)$ ({\it i.e.}, just saying that all the constituents move with the same velocity). The proper acceleration at $x^\mu (\zeta,\eta)$ is then identified as
 \begin{equation} \label{eq: aa-ao}
 \frac{1}{a(\zeta,\eta)} = \frac{1}{a_o(\eta)} + \zeta^1 \, .
 \end{equation}
\Eref{eq: aa-ao} implies limitation on the spatial extension of the system via the condition $1 + \zeta^1 a_o > 0$, since otherwise $x^\mu (\zeta,\eta)$ would enter into the Rindler horizon relative to the reference point.

Since the accelerations are point-dependent, so are also the proper-times. Combining \eref{eq: tauaeta} and \eref{eq: aa-ao} then yields
 \begin{equation} \label{eq: tatoeta}
 d\tau(\zeta,\eta) = d\tau_o(\eta) + \zeta^1 d\eta \, ,
 \end{equation}
so the relation between proper-time lapses and rapidity difference between any two simultaneity hyperplanes $\eta = \eta_1$ and $\eta = \eta_2$ is
 \begin{equation} \label{eq: DtauA}
 \Delta\tau \left(\zeta,\eta_1 \to \eta_2\right) = \Delta\tau_o \left(\eta_1 \to \eta_2\right) + \zeta^1 \Delta\eta \, .
 \end{equation}
($\Delta\eta = \eta_2 - \eta_1$). For any two points A and B then follows the relation between the proper-time lapses measured along their corresponding world-lines
 \begin{equation} \label{eq: DtauAB}
 \Delta\tau_{\rm B} \left(\eta_1 \to \eta_2\right) = \Delta\tau_{\rm A} \left(\eta_1 \to \eta_2\right) + \left(\zeta_{\rm B}^1 - \zeta_{\rm A}^1\right) \Delta\eta \, ,
 \end{equation}
independently of the choice of the reference point $x_o^\mu$ and the particular details of the acceleration. Recalling that rapidity differences and proper-times are Lorentz invariants verifies the Lorentz invariance of these results.

\vskip20pt

\section{Synchronization with light signals}\label{sec: synclight}

Spatially extended systems consist of a number of points moving on separate world-lines. Attaching a virtual clock to each point, proper-times may be measured at the points. If these clocks are synchronizable, proper-times measured at different points of the system may be linked and compared.

Synchronization may be understood here as ``adjusting the timing of two clocks''. When the clocks are separated, some mediating mechanism is required to link their time readings. As originally suggested by Einstein, this may be done by sending light signals, either from a common source situated in between to both clocks or by directly communicating light signals between them.

\begin{figure}
\includegraphics[width=9cm]{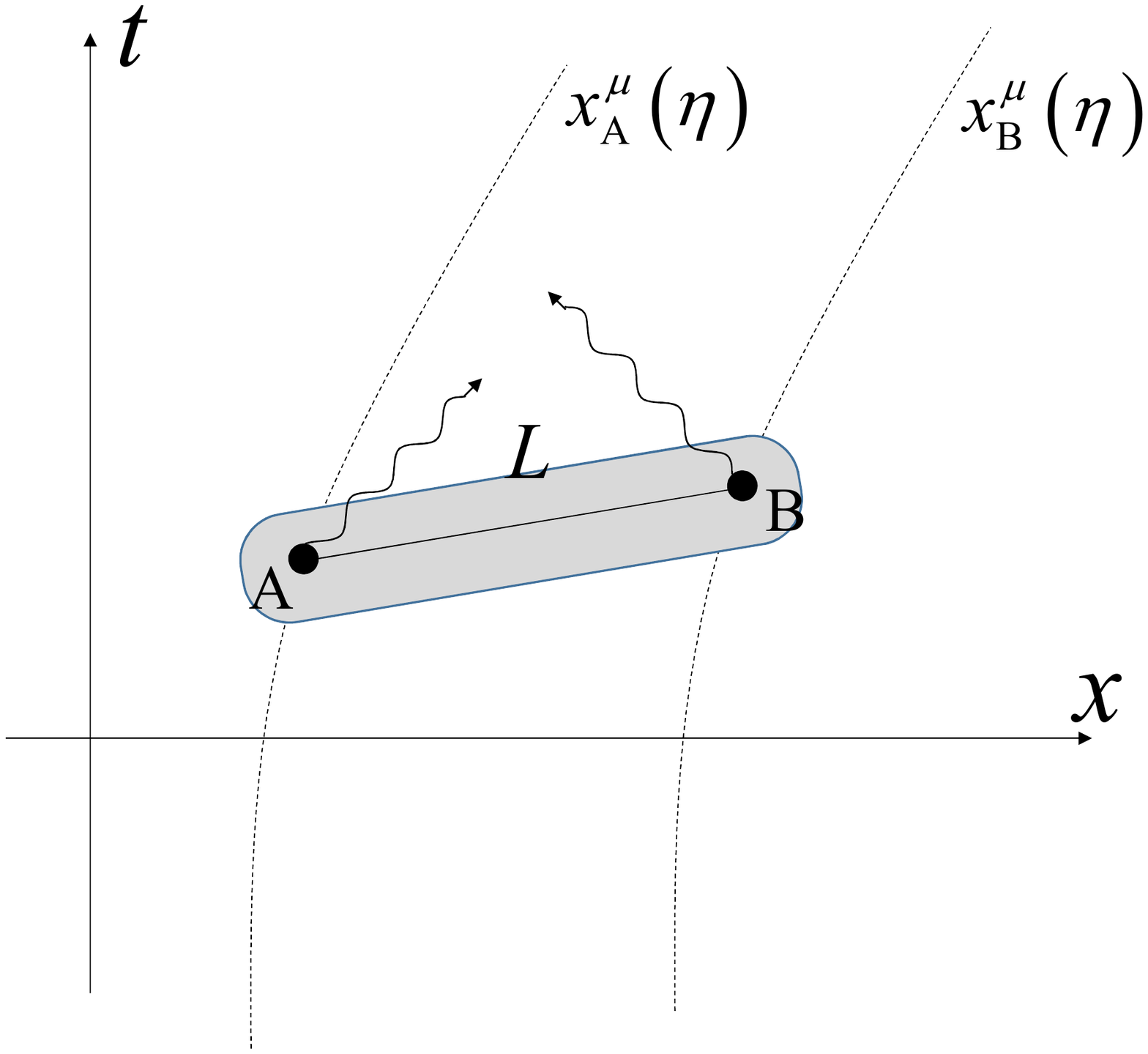}\\
\caption{Space-time diagram showing A \& B's world-lines, both A and B on a simultaneity hyperplane, and the signals emitted from both points towards each other.}\label{fig: signals}
\end{figure}

Let two clocks be located at points A, B along the $x$-axis in a rigidly accelerated system, with proper distance $L$, so that in the notation of \Sref{sec: rectirig} $\zeta_{\rm B}^1 - \zeta_{\rm A}^1 = L$. At a certain moment a light signal is simultaneously sent from A towards B and from B to A (\Fref{fig: signals}). If the system were inertial then it takes equal times for both signals to arrive to their destinations, allowing synchronization to be achieved and maintained. But if the system accelerates, time differences ensue. In the present section we discuss these differences and their dependence on the acceleration.

In terms of the rapidity the clocks' world-lines are given by \eref{eq: intx}
\begin{equation}\label{eq: xABeta}
 x_i^\mu(\eta) = \left(\int\limits_{}^\eta {\frac{\cosh\eta}{a_i(\eta)}d\eta},\int\limits_{}^\eta {\frac{\sinh\eta}{a_i(\eta)}d\eta},0,0\right) \quad i = {\rm A,B} \, ,
\end{equation}
satisfying $x_{\rm B}^\mu(\eta) = x_{\rm A}^\mu(\eta) + L n_1^\mu (\eta)$ with the proper accelerations related by \eref{eq: aa-ao},
 \begin{equation} \label{eq: aAaB-ao}
 \frac{1}{a_{\rm B}(\eta)} = \frac{1}{a_{\rm A}(\eta)} + L
 \end{equation}

Let us consider a signal emited from A at $\eta = \eta_1$ and arriving to B at $\eta = \eta_2$. The important quantity here is the rapidity difference $\Delta\eta = \eta_2 - \eta_1$ which is Lorentz invariant. The 4-vector displacement of the signal is $\Delta x_{\rm AB}^\mu = x_{\rm B}^\mu (\eta_2) - x_{\rm A}^\mu (\eta_1)$, and the light-cone condition on $\Delta x_{\rm AB}^\mu$ implies either
\begin{equation}\label{eq: lcondA}
 \int\limits_{\eta_1}^{\eta_2} {\frac{\cosh\eta}{a_{\rm A}(\eta)}d\eta} + L\sinh\eta_2 = \int\limits_{\eta_1}^{\eta_2} {\frac{\sinh\eta}{a_{\rm A}(\eta)}d\eta} + L\cosh\eta_2
\end{equation}
or, equivalently,
\begin{equation}\label{eq: lcondB}
 \int\limits_{\eta_1}^{\eta_2} {\frac{\cosh\eta}{a_{\rm B}(\eta)}d\eta} + L\sinh\eta_1 = \int\limits_{\eta_1}^{\eta_2} {\frac{\sinh\eta}{a_{\rm B}(\eta)}d\eta} + L\cosh\eta_1
\end{equation}
which yield
\begin{equation}\label{eq: LcondAB}
 L = \int\limits_{\eta_1}^{\eta_2} {\frac{e^{\left(\eta_2 - \eta\right)} d\eta}{a_{\rm A}(\eta)}} = \int\limits_{\eta_1}^{\eta_2} {\frac{e^{-\left(\eta - \eta_1\right)} d\eta}{a_{\rm B}(\eta)}}
\end{equation}
The light-cone condition \eref{eq: LcondAB} defines the relation between $L$, $\eta_1$ and $\eta_2$.

Let $\Delta \tau_{\rm A}\left(\eta_1,\eta_2\right),\Delta \tau_{\rm B}\left(\eta_1,\eta_2\right)$ be the proper-time lapses between signal emission and arrival as measured at A and B. From \eref{eq: tauaeta} it follows that
\begin{equation}\label{eq: Dtaui12}
 \Delta \tau_i\left(\eta_1,\eta_2\right) = \int\limits_{\eta_1}^{\eta_2} {\frac{d\eta}{a_i(\eta)}} \quad i = {\rm A,B}
\end{equation}
and from \eref{eq: aAaB-ao} it follows that
\begin{equation}\label{eq: LDtauAB}
 \Delta\tau_{\rm B} = \Delta\tau_{\rm A} + L\Delta\eta
\end{equation}
\Eref{eq: LcondAB} may now be used, in conjunction with \eref{eq: Dtaui12}, to get useful inequalities. First we get
\begin{equation}\label{eq: LDtauA}
 L = \int\limits_{\eta_1}^{\eta_2} {\frac{e^{\left(\eta_2 - \eta\right)} d\eta}{a_{\rm A}(\eta)}} > \int\limits_{\eta_1}^{\eta_2} {\frac{d\eta}{a_{\rm A}(\eta)}} = \Delta \tau_{\rm A}\left(\eta_1,\eta_2\right)
\end{equation}
and
\begin{equation}\label{eq: LDtauB}
 L = \int\limits_{\eta_1}^{\eta_2} {\frac{e^{-\left(\eta - \eta_1\right)} d\eta}{a_{\rm B}(\eta)}} < \int\limits_{\eta_1}^{\eta_2} {\frac{d\eta}{a_{\rm B}(\eta)}} = \Delta \tau_{\rm B}\left(\eta_1,\eta_2\right)
\end{equation}
Combining \eref{eq: LDtauA} and \eref{eq: LDtauB} yields the inequality
\begin{equation}\label{LDtauAB1}
 \Delta \tau_{\rm A}\left(\eta_1,\eta_2\right) < L < \Delta \tau_{\rm B}\left(\eta_1,\eta_2\right) \, ,
\end{equation}
so that the proper-time lapse for the signal transmission as measured at the emitter, $\Delta \tau_{\rm A}$, is shorter than $L$, while the corresponding lapse measured at the detector, $\Delta \tau_{\rm B}$, ia larger than $L$, in accordance with \eref{eq: LDtauAB}. Then, using \eref{eq: LDtauAB}, \eref{LDtauAB1} may be turned over to yield
\begin{equation}\label{LDtauAB21}
 (1 - \Delta\eta)L < \Delta \tau_{\rm A}\left(\eta_1,\eta_2\right) < L
\end{equation}
and
\begin{equation}\label{LDtauAB22}
 L < \Delta \tau_{\rm B}\left(\eta_1,\eta_2\right) < (1 + \Delta\eta)L
\end{equation}

If the signal is emited from B at $\eta = \eta_1$, arriving to A at $\eta = \eta_2$, then the light-cone condition \eref{eq: LcondAB} changes to
\begin{equation}\label{eq: LcondBA}
 L = \int\limits_{\eta_1}^{\eta_2} {\frac{e^{\left(\eta - \eta_1\right)} d\eta}{a_{\rm A}(\eta)}} = \int\limits_{\eta_1}^{\eta_2} {\frac{e^{-\left(\eta_2 - \eta\right)} d\eta}{a_{\rm B}(\eta)}} \, ,
\end{equation}
slightly different from \eref{eq: LcondAB} but leading to the same inequalities. In either case, these relations, applicable for arbitrary accelerations, are Lorentz invariant since rapidity differences are Lorentz invariant. The various proper-time differences $\Delta\tau$ depend on the details of the acceleration, but they are bounded by quantities that are acceleration-independent; it may therefore be appreciated that the dependence of the synchronization relations on the details of the acceleration is relatively limited.

As a specific example, for constant acceleration it follows from \eref{eq: LcondAB} and \eref{eq: LcondBA} that for signals in either direction (A $\to$ B or B $\to$ A)
\begin{equation}\label{eq: Deta.hyp}
 \Delta\eta = \ln \left( 1 + a_{\rm A}L \right) = - \ln \left( 1 - a_{\rm B}L \right)
\end{equation}
with the proper-time lapses
\begin{equation}\label{eq: DtauAB.hyp}
 \Delta\tau_{\rm A} = \frac{L \Delta\eta}{e^{\Delta\eta} - 1} \quad , \quad \Delta\tau_{\rm B} = \frac{L \Delta\eta}{1 - e^{-\Delta\eta}}
\end{equation}
Therefore, signals in both directions emitted simultaneously will also arrive simultaneously. This is not the case for varying acceleration.

These results are easily extended for signals emitted simultaneously from a common source situated in between A and B.

\vskip20pt

\section{Accelerating clocks}\label{sec: acclock}

The foregoing results bear consequences regarding the properties of accelerating clocks:

A clock is a device with an intrinsic periodic mechanism. It is convenient, whenever possible, to regard clocks ideally as point-like, because then their time evolution is confined to a single world-line. The idea of inertial point-like clocks is necessary for the construction of the Minkowski space-time continuum. Time is then assumed to be measured on non-inertial point-like clocks according to \eref{eq: Dtau}, independently of the details of the acceleration \cite{MTW73,Rindler2006}. The last assertion is known as the {\it clock hypothesis}.

So far there is no physical evidence to question the validity of the clock hypothesis, which was thus un-mentionally assumed at the beginning of the paper. Yet, this validity was questioned several times (see, {\it e.g.}, \cite{Mashhoon1990a,MainSted1993,Kowalski1996,Fletcher2013}). Our results add, theoretically, an argument in favour of the clock hypothesis : Real physical clocks are composite, spatially extended systems. Being regarded as composed of point-like constituents moving on different world-lines, these offer a continuum of proper-times within which the proper-time characterizing the clock must be identified. These proper-times may be related only when the clock is linearly-rigidly accelerating, and then the relation \eref{eq: DtauAB} between the proper-times is indeed independent of the details of the acceleration. In practice, even the largest difference between proper-time readings, which is $L\Delta\eta$, where $L$ is the spatial dimension of the clock, is very minute \cite{EJP2016a}, so the readings are very close : Let $L=10\rm{m}$, $v_1 = 0$ and $v_2 = 0.9c$. Then $\Delta\eta = \tanh^{-1}(0.9) = 1.47$ and $L\Delta\eta / c = 4.9\times 10^{-8} \rm{s}$. The discrepancy is real, but macroscopically hardly noticeable.

As is evident from the previous section, the details of the acceleration may enter only in the process of synchronization, or initial linking, of the time measurement along these world-lines. If the clock starts inertial and accelerates only after synchronization has been completed, then the consequent relations between the time-measurements are independent of the details of the acceleration. It is only when synchronization is attempted while the clock is already accelerating that time-measurements depend on the acceleration, but here, again, the discrepancies are minute, of the order of $L \Delta\eta$.

\vskip20pt

\section{The ambiguity of proper-time comparison in non-rigid motion}\label{sec: nonrig}

The proper-time lapse computed relative to an external inertial observer is given by \eref{eq: Dtau}. This proper-time lapse is Lorentz-invariant when computed between two fixed events ({\it i.e.}, independent of the external observer), but if the limiting events are determined by the external observer the situation changes, even for inertial motion.

\begin{figure}
\includegraphics[width=9cm]{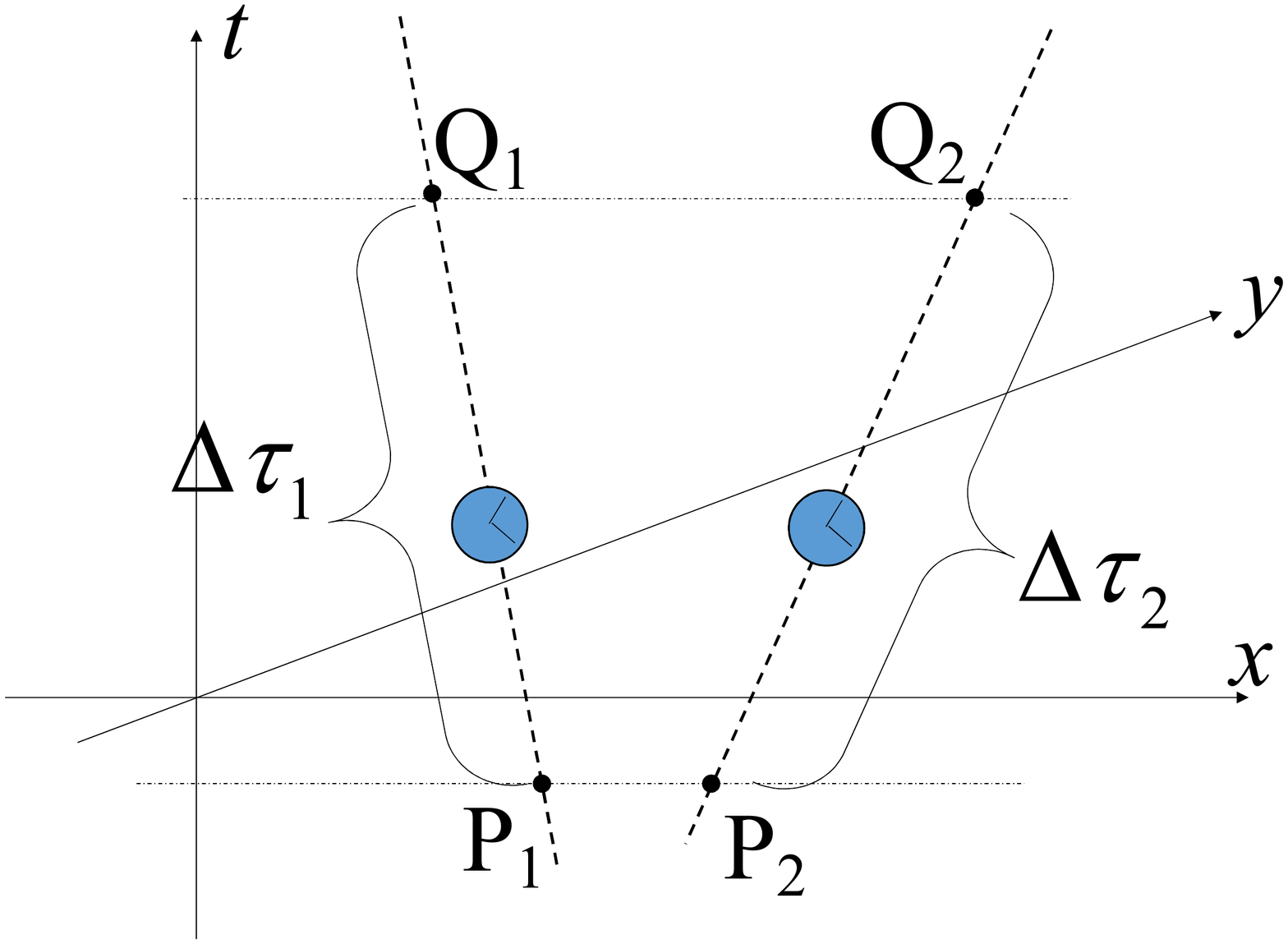}\\
\caption{Space-time diagram showing the world-lines of two relatively moving inertial point-like clocks. Proper-time lapses are measured or computed between simultaneity hyperplanes $\rm{P}_1\rm{P}_2$ and $\rm{Q}_1\rm{Q}_2$ relative to the external observer.}\label{fig: 2inertial}
\end{figure}

Consider two point-like clocks moving with different constant velocities (\Fref{fig: 2inertial}). The relation between the proper-time lapses, measured along their world-lines between events simultaneous relative to the same external observer, are
\begin{equation}\label{eq: zzz}
 \frac{\Delta\tau_1}{\Delta\tau_2} = \frac{\sqrt{1 - {v_1}^2}}{\sqrt{1 - {v_2}^2}}
\end{equation}
Using the rapidities for both velocities, with $\eta_{1,2} = \eta\left(v_{1,2}\right)$ and $\Delta\eta = \eta_2 - \eta_1$,
\begin{eqnarray}\label{eq: yyy}
 \frac{\Delta\tau_1}{\Delta\tau_2} &=& \frac{\cosh\eta_2}{\cosh\eta_1} = \frac{\cosh\left(\eta_1 + \Delta\eta\right)}{\cosh\eta_1} = \cosh\left(\Delta\eta\right) + \frac{\sinh\eta_1}{\cosh\eta_1} \sinh\left(\Delta\eta\right) = \nonumber \\
 &=& \cosh\left(\Delta\eta\right) + v_1 \sinh\left(\Delta\eta\right)
\end{eqnarray}
The rapidity difference $\Delta\eta$ is Lorentz-invariant, but $v_1$ is observer-dependent. Since the range of possible $v_1$ values is $-1 < v_1 < 1$, then follows the inequality
\begin{equation}\label{eq: Dtau12}
 e^{- \Delta \eta} < \frac{\Delta\tau_1}{\Delta\tau_2} < e^{\Delta \eta}
\end{equation}
which defines the measure of ambiguity of the proper-times ratio, which is clearly observer-dependent. The inequality \eref{eq: Dtau12} is Lorentz invariant due to the Lorentz invariance of the rapidity difference $\Delta\eta$.

\vskip20pt

\section{The case of two spaceships -- an example for the correct usage of simultaneity}\label{sec: 2sp}

As has already been pointed out, comparison of proper-time lapses requires introducing proper simultaneity; and since simultaneity is frame dependent and not preserved by Lorentz transformations, much care has to be taken at this point. In the present section we discuss an example that stresses and highlights the need for correct choice of the simultaneity hyperplanes :

Two small (so that they may be regarded point-like) spaceships, A and B, are at rest in the (inertial) home-base, distance $L$ apart. At a certain moment a signal is received causing the engines of the spaceships to ignite and they both simultaneously embark into space following what seems to be the same journey plan relative to the home-base reference frame $S_{\rm H}\left\{\left(x^\mu\right)\right\}$
 \begin{equation} \label{eq: xAxB.B}
 x_{\rm A}^\mu = \left(t_{\rm A},\xi(t_{\rm A}),0,0 \right) \qquad , \qquad  x_{\rm B}^\mu = \left(t_{\rm B},\xi(t_{\rm B}) + L,0,0 \right)
\end{equation}
where $\xi(t)$ describes some non-uniform motion.

Two questions may be asked :
\begin{enumerate}
 \item {A stretchable string connects the two spaceships. Will the string stretch and eventually be torn apart ?}
 \item {The astronauts in both spaceships are twins. Will they still be of the same age after the journey ?}
\end{enumerate}
Both questions have been raised up in the literature and discussed to some extent. Question (i) is the basis for the so-called `Bell's spaceships paradox' \cite{BellSSP}. Question (ii) was raised by Boughn \cite{Boughn1989}, referring to the astronauts as `identically accelerated twins'. Both questions require comparison, either for the relative distance or for the proper-time difference, between the spaceships. Comparison, as we have already seen, requires a common platform, which must be a proper simultaneity hyperplane. The quality of the answers depends, therefore, on the correct choice of the reference frames which define the correct simultaneity hyperplanes :

The scenario \eref{eq: xAxB.B} requires that the relative distance between the spaceships remain constant relative to the home-base reference frame during the journey; but this simultaneity is irrelevant -- the astronauts can measure time, distances, velocities and accelerations only relative to themselves. If the spaceships are connected by a string, any information regarding the state of its stretching can only be obtained by direct measurement relative to the spaceships themselves.

Therefore, the concept of {\it proper distance} must be introduced -- the distance between two points which are simultaneously at rest. Without requiring proper simultaneity the whole concept of length or distance between two world lines is completely ambiguous.

The need for proper simultaneity is clear when we consider that the spaceships need to operate their engines in order to accelerate. The engines are operated by the computers of the spaceships which need to be programmed accordingly. The operation program necessarily uses time for the different stages of the journey. Whose time is it ? it can only be the proper-time of the spaceship, which is the time provided by the spaceship's clock (which, quite reasonably, constitutes a component of the computer). The accelerations of the spaceships are therefore determined relative to their own proper-times, and may be compared unambiguously only on proper simultaneity hyperplanes. Therefore, the twins in \eref{eq: xAxB.B} are not really {\it identically} accelerated.

Therefore, for questions that concern proper distances and proper times the correct simultaneity hyperplanes must be relative to the spaceships themselves, {\it i.e.}, in which both spaceships are momentarily at rest.

Moving so that there is always a common momentary rest frame for both spaceships, implies that the motion must be rigid, with world-lines given by \eref{eq: xABeta}. The equal-$\eta$ simultaneity hyperplanes correspond to the common motion of the two spaceships at the string's ends. On the other hand, the motion proposed in \eref{eq: xAxB.B} is certainly non-rigid, since constant relative distance is assumed relative to the home-base rather than between the spaceships.

It is easy to demonstrate explicitly that there is no common rest frame (therefore no common simultaneity hyperplane) in Bell's and Boughn's scenario \eref{eq: xAxB.B} if the spaceships accelerate : It suffices to assume hyperbolic motion (constant proper acceleration). Writing the home-base scenario \eref{eq: xAxB.B} in terms of the rapidities $\eta_{\rm A,B}$,
 \begin{equation} \label{eq: xAxB.hyp}
 \fl \hskip40pt x_{\rm A}^\mu = \left(\rho\sinh\eta_{\rm A},\rho\cosh\eta_{\rm A},0,0 \right) \quad , \quad  x_{\rm B}^\mu = \left(\rho\sinh\eta_{\rm B},\rho\cosh\eta_{\rm B} + L,0,0 \right)
\end{equation}
($a = \rho^{-1}$ is the common proper acceleration), and taking into account that the momentary velocities are $v_{\rm A,B} = \tanh\eta_{\rm A,B}$, the spaceships have a common rest frame at the home-base only for $\eta_{\rm A} = \eta_{\rm B} = 0$, just before launching into their journey. Lorentz transforming to another inertial reference frame $S_{\rm R}\left\{\left(\bar x^\mu\right)\right\}$ moving with velocity $V_{\rm R} = \tanh\eta_{\rm R}$ relative to the home-base, the world-lines \eref{eq: xAxB.hyp} become
\begin{eqnarray} \label{eq: xAxB.hypR}
 \fl \hskip40pt \bar x_{\rm A}^\mu &=& \left(\rho\sinh\left(\eta_{\rm A} - \eta_{\rm R}\right),\rho\cosh\left(\eta_{\rm A} - \eta_{\rm R}\right),0,0 \right) \, , \nonumber \\
 \fl \hskip40pt \bar x_{\rm B}^\mu &=& \left(\rho\sinh\left(\eta_{\rm B} - \eta_{\rm R}\right) - L\sinh\eta_{\rm R},\rho\cosh\left(\eta_{\rm B} - \eta_{\rm R}\right) + L\cosh\eta_{\rm R},0,0 \right)
\end{eqnarray}
$\bar\eta_{\rm A,B} = \eta_{\rm A,B} - \eta_{\rm R}$ are the rapidities relative to $S_{\rm R}$. Both spaceships are momentarily at rest in $S_{\rm R}$ when $\eta_{\rm A} = \eta_{\rm B} = \eta_{\rm R}$, but these two events are not simultaneous -- they correspond to different $S_{\rm R}$-times -- $\bar x_{\rm A}^0 = 0$ and $\bar x_{\rm B}^0 = - L\sinh\eta_{\rm R}$. Consequently, if the spaceships accelerate, it is impossible to find in any inertial reference frame two simultaneous events in which both space-ships are momentarily together at rest. Simultaneity could be achieved only with rigid motion, with
\begin{eqnarray}\label{eq: xAxB.hyprig}
 x_{\rm A}^\mu &=& \left(\rho\sinh\eta_{\rm A},\rho\cosh\eta_{\rm A},0,0 \right) \quad , \nonumber \\
x_{\rm B}^\mu &=& \left(\left(\rho + L\right)\sinh\eta_{\rm B},\left(\rho + L\right)\cosh\eta_{\rm B},0,0 \right)
\end{eqnarray}
which coinsides with \eref{eq: xAxB.hyp} only at the home-base rest-frame, with $\eta_{\rm A} = \eta_{\rm B} = 0$.

Therefore, when the motion is not rigid the whole problem is ill-posed right from the start : To be able to measure the string's proper length it must be at rest, even momentarily, relative to some inertial reference frame which defines a simultaneity hyperplane; but in a non-rigid motion the two spaceships do not share any common simultaneity hyperplane. Therefore, the mere concept of proper length is meaningless in non-rigid motion. Similarly, for world-lines in non-rigid motion there is no way to even compare the ages.

Both questions above are answerable {\it only} if the motion is rigid. The answers are then immediate :
\begin{enumerate}
  \item {The string maintains constant proper length, with differential acceleration along it. The components of the string feel stresses, but these are only required to maintain the accelerated rigid motion and they don't change the string's length.}
  \item {The astronauts' ages are determined by the proper-time lapses along their world-lines. Comparison of the ages requires simultaneity hyperplanes, which exist only for rigid motion. Then it follows from \eref{eq: LDtauAB} that the ages differ by $L \Delta\eta$, {\it i.e.}, depending on the proper distance between the twins and the rapidity difference between the home-base and the end station.}
\end{enumerate}

The two questions coincide with a recent discussion of Bell's `paradox' by Franklin \cite{Franklin2010}, who, among other things, also compared the Minkowskian times of the right and left spaceships (or brothers) which are obviously the same in any instantaneous rest frame. However, the {\it ages} of the brothers are determined not by the Minkowskian times but by the proper-times measured along their (separate) space-time trajectories. If the end station moves relative to the home-base (so that $\Delta\eta \ne 0$ between the initial and final states) then the brothers do indeed end up with different ages, simply because of siting in separated spaceships.

\vskip20pt

\section{Concluding remarks}\label{sec: conrem}

The ages or proper-times measured at different constituents of an extended system may be related and compared only if momentary simultaneity hyperplanes may be identified along the system's journey in space-time. The relation of proper-time lapses at two distinct points of an accelerating system is then uniquely determined, Lorentz covariantly, only for rectilinear relativistic rigid motion. Rectilinear rigid motion may therefore serve to model comparative proper-time measurement in accelerated relativistic systems. This modelling was used here to reflect upon and discuss the clock hypothesis and the correct use of simultaneity.

Besides being the characterizing property that allows proper-time comparison, rigidity has a value of its own also for the following reason : The space-time picture of extended systems is of a congruence of world-lines. What makes this congruence a ``system", more than just only a collection of world-lines ? Such a collection becomes a ``system" -- a whole that is more than just the sum of its constituents -- when there is a property which does not pertain to the individual constituents but characterizes the group as a whole. Rigidity is such a property.

In addition, we point out the useful use of the rapidity $\eta$ as the parameter of evolution for linearly accelerated systems.

Finally, we also make note of the fact that while the continuum picture of Minkowski space-time uses point-like clocks, these are only idealizations of real clocks which are necessarily spatially-extended. We may therefore conclude that the continuum picture of Minkowski space-time is only approximately self-consistent, even without taking into account gravitation and quantum mechanics.

\vskip20pt

\rule{10cm}{1pt}


 \end{document}